\documentclass[twocolumn,preprintnumbers,amsmath,amssymb]{revtex4}

\usepackage{graphicx}  
\usepackage{sidecap}
\begin{document}

\title{Interband infrared photodetectors based on HgTe--CdHgTe  quantum-well heterostructures
}
\author{V. Ya. Aleshkin$^1$, A. A. Dubinov$^1$,  S. V. Morozov$^1$, M. Ryzhii$^2$,  T.~Otsuji$^3$, V. Mitin$^{4}$,\\ 
M. S. Shur$^5$, and V. Ryzhii$^{3,6,7}$ 
}
\address{
$^1$Institute for Physics of Microstructures of RAS and Lobachevsky University of Nizhny Novgorod,
Nizhny Novgorod, 603950, Russia\\
$^2$Department of Computer Science and Engineering, University of Aizu, Aizu-Wakamatsu 965-8580, Japan\\
$^3$Research Institute of Electrical Communication, Tohoku University, Sendai 980-8577, Japan\\
$^4$ Department of Electrical Engineering, University at Buffalo, Buffalo, New York 1460-1920, USA\\
$^5$ Department of Electrical, Computer, and Systems Engineering, Rensselaer Polytechnic Institute, Troy, New York 12180,
$^6$Institute of Ultra High Frequency Semiconductor Electronics of RAS,\\
 Moscow 117105, Russia\\
$^7$Center for Photonics and Infrared Engineering, Bauman Moscow State Technical University, Moscow 111005, 
}

 \begin{abstract} 
\noindent{\bf Keywords:} quantum well, HgTe-CdHgTe heterostructure,  infrared photodetector, responsivity, detectivity.\\ 
We calculate the characteristics of  interband HgTe-CdHgTe  quantum-well infrared photodetectors (QWIPs).
Due to a small probability of the electron capture into the QWs, the interband HgTe-CdHgTe QWIPs can exhibit very high photoconductive gain.  Our analysis demonstrates  significant potential advantages of these devices compared to the conventional CdHgTe photodetectors and the  A$_3$B$_5$ heterostructures.
\end{abstract} 

\maketitle

\newpage

\section{Introduction}
 \vspace{-0.2cm}

\begin{SCfigure*}
\centering
\includegraphics[width=10.0cm]{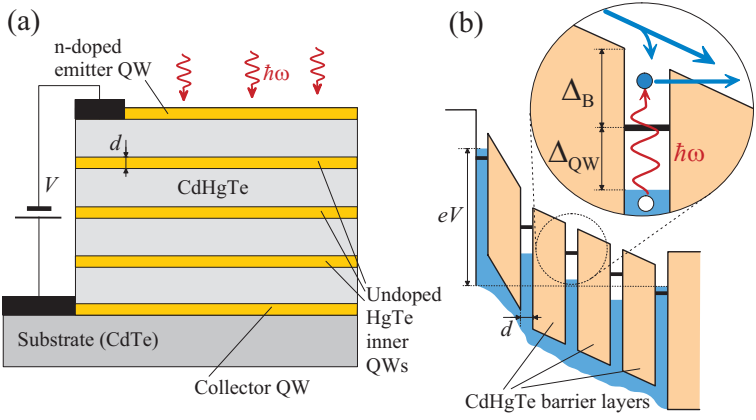}
\caption{Schematic views of (a)  interband HgTe-CdHgTe QWIP structure, (b) its band diagram at the bias voltage $V$. The inset in Fig.1(b)
shows a fragment of the    band diagram in more detail.
Wavy arrows correspond to the incident photons and to the processes of the electron transitions from the subband of the QW valence band  to the electron subband (above its bottom). Solid  arrows indicate the  propagation of the  electrons  (both injected from the emitter and the preceding QWs) above the barriers, capture of these electrons into the QW, and tunneling of the photoexcited electrons from the QW.}
\end{SCfigure*}

The intersubband quantum-well infrared photodetectors (QWIPs)~\cite{1,2,3,4} based on the A$_3$B$_5$ heterostructures  have been developed since 1960s~\cite{5,6,7,8}. These photodetectors  are still the subject of intensive theoretical
and experimental studies~\cite{9,10,11,12,13,14,15}.
Using different  materials  for the QWs and the inter-QW barrier layers, and varying the QW width,
 one
can adjust the QWIP operation  from near-infrared to terahertz frequencies. 
The main advantage of QWIPs is associated with the maturity of the material and processing  technology allowing to produce the QWIPs and QWIP-based large arrays with desirable 
spectral characteristics, including multicolor systems. Another advantage is the ease of integration with other devices. Due to the mature technology, the QWIP-based large arrays exhibit 
stability and  high pixel-to-pixel uniformity.  
The cost of the A$_3$B$_5$ QWIP based
devices and systems is much less than of those based on CdHgTe and InSb. 
The intersubband QWIP disadvantages
include 
 the large thermal dark current  at elevated temperatures, which prevents the room operation~\cite{1,2,3}
and the need  
for the radiation coupling structures (for the n-type QWIPs). 
Therefore, the standard A$_3$B$_5$ QWIPs cannot compete with the interband CdHgTe photodiodes (PDs)~\cite{3} in performance. 
The recent progress in the fabrication of CdHgTe QW heterostructures~\cite{16,17,18,19} provides an opportunity
for a further enhancement of the  CdHgTe photodetector technology.
In this paper, we propose and evaluate the QWIPs using the interband transitions in the CdHgTe heterostructures, in particular,  with the  HgTe QWs. 
The interband HgTe-CdHgTe  QWIP  operation is associated with the photoexcitation of the electrons in the QWs  followed by their escape. 
These processes result in the redistribution of the device potential,   varying
 the electric field at the emitter and  the electron injection current from
the emitter. The electric potential distribution in question is governed by the balance of the photoexcitation from the QWs and the electron capture into the QWs similar to what takes place in the standard QWIPs~\cite{20,21,22,23,24}.
Due to the features of the energy spectrum in the QWs, these  interband QWIPs can operate at the normal IR radiation incidence. In this regard, the interband HgTe-CdHgTe QWIP operation is akin to the operation of the vertical graphene-layer infrared photodetectors (GLIPs)~\cite{25,26,27}.

The interband electron transitions in the HgTe-CdHgTe QWIPs can provide a rather strong absorption.
A  weak capture of the electrons into the  QWs 
leads to a high photoconductive gain (phototransistor effect).
The required spectral characteristics of the interband QWIPs can be obtained by a proper choice of the CdHgTe composition and the QW width. The possibility of the QW engineering enables the fabrication of the interband HgTe-CdHgTe QWIPs with desirable spectral
characteristics (by using, for example, the HgTe QWs with different width). 
We compare the  QWIPs under consideration  with  the standard p-i-n CdHgTe   photodiodes (PDs) 
and discuss the advantages of the former.

\section{Interband QWIP device model}

The QWIP structure under consideration consists of a number of the undoped HgTe QWs   ($N = 1,2,3,...$) 
with the energy gap, $\Delta_{QW} $,  between the top of the highest hole subband and the bottom of the lowest electron subband. The QWs are  
separated by a material with the  energy gap $\Delta_G > \Delta_{QW}$  ( Cd$_x$Hg$_{1-x}$Te with $0 < x < 1$),
so that the barrier  of the height $\Delta_B$ is formed for electrons (see Fig.~1).
The structure is sandwiched between the emitter and collector n-doped layers. For the definiteness we assume
that both these layers are the same QWs (as the inner QWs) but highly doped by donors.
Figures ~1(a) and 1(b) show the QWIP device composition and the band diagram under sufficiently strong bias voltage $V \gg V_{bi}, k_BT/e$
(where $V_{bi}$ is the built-in voltage between the n-doped contact and the undoped inner QWs, $T$ is the temperature, $k_B$ is the Boltzmann constant, and $e$ is the electron charge).

The electron photoexcitation in the QWs in question under the normally incident radiation  can be associated with both the intersubband
transitions within the QW conduction band and with the interband transitions.
Considering that the electron density in the undoped QWs is relatively small and that the pertinent matrix elements
are  small, we focus on the contribution of the interband transitions.

To provide an effective escape of the photoexcited electrons from the QWs into the states above the barriers,
the following conditions are assumed:

\begin{equation}\label{eq1}
\hbar\omega \gtrsim \Delta_{QW} + \biggl(1 + \frac{m}{M}\biggr)\Delta_B \simeq \Delta_{QW}+ \Delta_B = \hbar\omega_{th}.
\end{equation}
Here $m$ and $M$ are the electron and hole effective masses in the QWs, $\Delta_{B}$ is the energy separation
between the bottom of the barrier conduction band and the bottom of the lowest electron  subband in the QW
(the barrier height for the electrons in the QWs),
$\hbar\omega_{th}$ is the threshold of the effective photoexcitation. At $\hbar\omega > \hbar\omega_{th}$,
the electrons photoexcited in the QW can easily escape. In the opposite case,
the electron escape from the QWs is associated with the tunneling through the triangular barrier  formed by
the electric field.

Using a simplified model for the characteristics of the vertical photodetectors using the photoexcitation from
the localized states in the structure and the electron injection from the emitter (used previously in the papers on the standard QWIPs as well as in the GLIPs~\cite{19,20,21,22,23,24,25,26,27}), one can obtain for the photocurrent density in the QWIP $j_{photo}$ and
its photodetector responsivity $R = j_{photo}/I_{\omega}\hbar\omega$

\begin{equation}\label{eq2}
j_{photo}  = \frac{e\beta_{\omega}\theta_{\omega}\xi_{N}}{p_c}I_{\omega}, \qquad  R= \frac{e\beta_{\omega}\theta_{\omega}\xi_{N}}{p_{c}\hbar\omega}.
\end{equation}
Here $\beta_{\omega}$  and $p_c$ are  the probabilities of the interband electron photoexcitation between the hole and electron subbands in the QW (radiation absorption coefficient)  and the capture into the QW, $\theta_{\omega}$ is the probability of the escape of the photoexcited electrons from the QW, $I_{\omega}$ and $\hbar\omega$
are the incident radiation flux and the photon energy, respectively.

The factor $\xi_N \lesssim 1$ describes a nonideality of the emitter (see Ref.~\cite{26} and the references therein): $\xi_N \simeq N/(\gamma^{3/2} + N)$ with $\gamma = (\Delta_B -\varepsilon_F)/\Delta_B$, where 
$N$ is the number of the inner QWs, $\varepsilon_F$ is the electron Fermi energy in the emitter QW counted from bottom of the lowest electron subband in the QW
and $\Delta_B$ is
 the energy spacing between the barrier top and the bottom
of this  subband  [see Fig.~1(b)].
Equation~(2) corresponds to the net rate of the photoescape $\beta_{\omega}\theta_{\omega}\xi_{N}N$ and
the phoconductive gain $g = 1/(Np_c)$.

At $\hbar\omega \leq \hbar\omega_{th}$ and $\hbar\omega \geq \hbar\omega_{th}$,

\begin{equation}\label{eq3}
\theta_{\omega} = \biggl\{1 +  \displaystyle\frac{\tau_{esc}}{\tau_{relax}}
\exp\biggl[\biggl(\frac{\omega_{th} -\omega}{\omega_{th}}\biggr)^{3/2}
\frac{E_{tunn}}{E}\biggr]\biggr\}^{-1}, 
\end{equation}

\begin{equation}\label{eq4}
\theta_{\omega} = \biggl(1 +  \displaystyle\frac{\tau_{esc}}{\tau_{relax}}\biggr)^{-1}, 
\end{equation}
respectively, where $\tau_{esc}$ and $\tau_{relax}$ are the try-to-escape and energy relaxation times, 
$E_{tunn} = 4\sqrt{2m_B}(\hbar\omega_{th}^{3/2})/3e\hbar$ and $E$ are the characteristic tunneling field (see, for example, Ref.~\cite{28}) and the electric field in the inter-QW barriers, respectively, and $m_B$ is the electron effective mass in the barrier material.

\section{Energy spectra, spectral characteristics of the interband absorption  and capture probabilities}

The functions $\beta_{\omega}$ and $\theta_{\omega}$ depend on the QW energy spectra,
which, in turn, depend on the compositions of the QWs and barrier layers  materials  and on the QW width, $d$. We calculated the energy  spectra and the spectral characteristics of the absorption coefficients for the heterostructures  grown on [013] CdTe surface with different content of Cd in the barrier layers at  temperatures $T = 77$ and 200~K.
 The refractive index of the barrier is set to be $n = \sqrt{15.2}$.

Figure~2 shows examples of the energy spectra of  the HgTe QWs   
surrounded by the Cd$_{0.27}$Hg$_{0.73}$Te and Cd$_{0.3}$Hg$_{0.7}$Te barriers calculated for different values of the QW width  and $T = 77$~K
and $T = 200$~K.  The spectra shown in Fig.~2 correspond to  two lowest electron subbands
marked as e$1$ and e$2$   
(virtually undistinguished due to a weak   interface inversion asymmetry splitting)  and two sets of split hole subbands marked as 
h$1$, h$2$  and h$3$, h$4$.

Figure~3 shows the energy gap, $\Delta_{QW}$, between the hole and electron subbands and the energy separation, $\Delta_B$, between the barrier conduction band and the bottom of the lowest electron subband (the barrier height)  as functions of the QW width $d$
calculated for different content of Cd "$x$"  in the barriers at different temperatures. 

\begin{figure}[t]
\centering
\includegraphics[width=8.5cm]{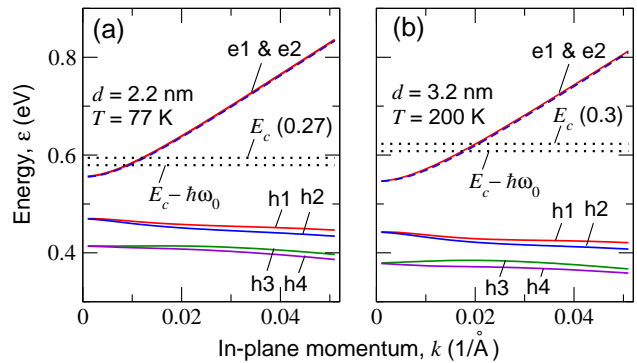}
\caption{Energy spectra of the HgTe QWs (a)  with the width $d = 2.2$~nm surrounded by the Cd$_{0.27}$Hg$_{0.73}$Te barriers at $T = 77$~K and (b) with the width $d = 3.2$~nm surrounded by Cd$_{0.3}$Hg$_{0.7}$Te barriers at $T = 200$~K: e1 and e2 lines
correspond so slightly split lowest electron subband, h1, h2 and h3,h4 correspond to two split hole subbands. Horizontal dotted lines show the bottom of the barrier conduction band $E_c$
(for the Cd contents $x = 0.27$ and $x=0.3$) and 
  the conduction band  bottom energy minus the optical phonon energy $\hbar\omega_0$, respectively. The energy is counted from the top of the CdTe valence band.}
\end{figure}

The calculations of $\beta_{\omega}$ are based on the following equation:

\begin{eqnarray}\label{eq5}
\beta_{\omega} = \frac{e^2}{c\omega\,n}\int\int d^2{\bf k}\sum_{i,j}
\frac{|\nu_{i,j}^x|^2 +  |\nu_{i,j}^y|^2  }{2\pi}\nonumber\\
\times\frac{\Gamma}{[(\varepsilon_i({\bf k}) - \varepsilon_j({\bf k}) + \hbar\omega)^2 + \Gamma^2]}.
\end{eqnarray}
Here $\Gamma $ characterizes broadening of the energy spectrum, the indices $i$ and
$j$ correspond to the initial and final states, $\nu_{i,j}^x$, $\nu_{i,j}^y$ are the matrix elements of the velocity operator, calculated in the framework of the $8\times 8$ Kane model
(see, for example,~\cite{29}), and $c$ is the speed of light. The spectra broadening is set to be $\Gamma =1$~meV.

In the calculations of the capture probability, $p_c$, 
we assumed that the capture of the electrons propagating over the inter-QW barriers into the QWs is primarily associated  with the emission of optical phonons (with the energy $\hbar\omega_0 = 0.015$~eV).

\begin{figure}[t]
\centering
\includegraphics[width=8.5cm]{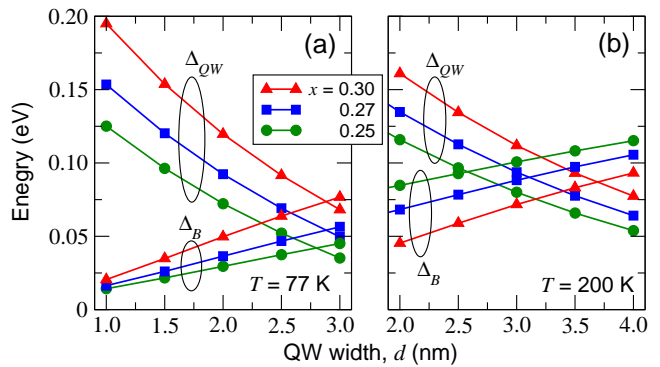}
\caption{The energy gap $\Delta_{QW}$ and the barrier height $\Delta_B$ versus the QW width $d$ for different 
Cd content $x$ at (a) $T = 77$~K and (b)  $T = 200$~K.}
\end{figure}

\begin{figure}[t]
\centering
\includegraphics[width=8.5cm]{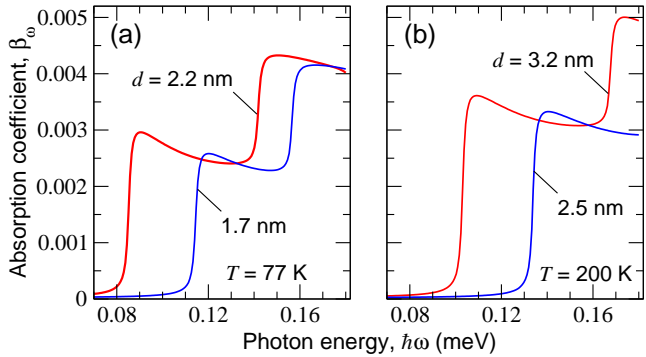}
\caption{The interband absorption  coefficient $\beta_{\omega} $
versus photon energy $\hbar\omega$ (a)  in the HgTe--Cd$_{0.23}$Hg$_{0.73}$Te heterostructures  with the QW widths  $d = 2.2$~nm and $d = 1.7$~nm (at  $T = 77$~K)   and at (b) in the HgTe--Cd$_{0.3}$Hg$_{0.7}$Te heterostructures
 with the QW thicknesses $d = 3.2$~nm   and   $d = 2.5$~nm (at $T = 200$~K). }
\end{figure}

Figure~4 shows the spectral characteristics of  the absorption coefficient (absorption probability) associated with the interband transitions in the HgTe--Cd$_{0.27}$Hg$_{0.73}$Te  (at $T = 77$~K) and in the HgTe--Cd$_{0.3}$Hg$_{0.7}$Te (at $T =200$~K) heterostructures with the HgTe QWs of different width $d$. The two-step increase in the absorption probability as a function of the photon energy is associated with  a noticeable split  of the h1, h2 and h3, h4 hole subband pairs (see Fig.2).

\section{Interband QWIP characteristics}

Figures~5 and 6 show the spectral characteristics  of the responsivity  $R$ for the interband HgTe-CdHgTe QWIPs with the same parameters as those in Fig.~4 and
$\xi_N = 0.738$ ($\gamma = 0.5$ and $N = 1$) calculated using Eqs.~(2) - (4) for different relative electric fields $U = E/E_{tunn}$. It is assumed that $\tau_{esc}/\tau_{relax} = 0.1$.

The plots in Fig.~5(a) correspond  to the capture probabilities   $p_c = 2.5\%$ and $p_c =1.1\%$ of the electrons with the average energy in the barrier layers  $\varepsilon = k_BT \simeq 6 $~meV.  The plots in Fig.~5(b) correspond to  $p_c = 0.36\%$ and $p_c = 0.58\%$
(of the average electron energy $\varepsilon = k_BT \simeq 17 $~meV).

The obtained values of the capture probability $p_c \simeq (0.36- 0.58)\% $ 
are generally of the same order of magnitude (or somewhat larger) as the  $p_c$ values  for the GLs~\cite{30}. 
The fact that $p_c$ in the QWIP under consideration  can be  larger at some QW parameters than in GLs,  can be explained by large energies of the optical phonons in GLs in comparison with HgTe or CdHgTe. As a consequence, the emission of the optical phonons accompanying  the electron capture into GLs requires larger variations of the electron momentum.

As follows from Eq.~(2), the dependence of the interband QWIP responsivity  on the number of the inner QWs,  $N$, is determined by the factor $\xi_N = N/(\gamma_N^{3/2} + N)$ with
 $\gamma = (\Delta_B - \varepsilon_F)/\Delta_B)$, where $\varepsilon_F$ is the Fermi energy in the emitter QW counted from the electron subband bottom. Since $\gamma$ varies from unity ($\varepsilon_F \simeq 0$) to zero ($\varepsilon_F = \Delta$  for a heavily doped emitter QW),
the factor $0.5 < \gamma_N <1$  for arbitrary $N$. This implies that the interband QWIP responsivity is a weak function of the number of inner QWs. A similar situation occurs
in the intersubband QWIPs and interband GLIPs.

Some  increase  (although a relatively slow) in  the interband QWIP responsivity with increasing   number of the QWs $N$ is confirmed
by Fig.~6 ($N = 1, 3,$ and 10).  At $N \gtrsim 10$, an increase in $R$ with $N$ becomes insignificant.

Larger  values of the responsivity corresponding to   the plots in Fig.~5(a) [and in Fig.~6(a)] in comparison to those shown in Fig.~5(b) [and in Fig.~6(b)] are attributed to a smaller value
of the capture probability $p_c$ for the former figure.

As follows from the above data, the interband QWIP responsivity  can be fairly large.

\begin{figure}[t]
\centering
\includegraphics[width=8.5cm]{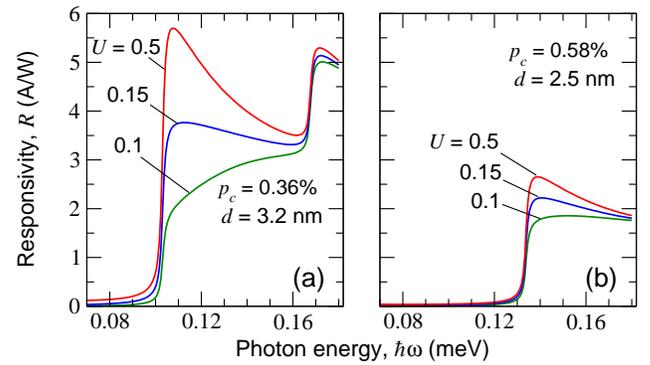}
\caption{ The spectral dependences of the responsivity $R$
for the interband  HgTe--Cd$_{0.3}$Hg$_{0.7}$Te QWIPs with $\gamma = 0.5$, $N = 1$,  and different normalized electric fields $U = E/E_{tunn}$:
(a)  $d = 3.2$~nm ($p_c = 0.36\%$) and (b) $d = 2.5$~nm ($p_c =0.58\%$)
at $T = 200$~K.}
\end{figure}

\section{Comparison with other photodetectors}

The interband HgTe-CdHgTe QWIPs surpass the QWIPs  using the intersubband   (intraband) both based on similar HgTe-CdHgTe
heterostructures and on the  A$_3$B$_5$ QW heterostructures. This is due to higher probability of the interband 
electron photoexcitation in the former compared to the probability of the intersubband photoexcitation.
The use of the intersubband QWIPs requires a substantial donor doping of the  QWs. However the latter leads to a decrease in the thermogeneration activation energy and  negatively affects the detector detectivity.

Below  we compare the responsivity and dark-current-limited detectivity of the interband QWIPs under consideration with  the traditional interband  p-i-n PDs. 
The operation of the latter is associated with the interband
photogeneration of the electrons and holes  in the depleted bulk i-layer CdHgTe
and their propagating in the vertical direction.

\vspace{-0.4cm}

\subsection{Responsivity}

\vspace{-0.4cm}

\begin{figure}[t]
\centering
\includegraphics[width=8.5cm]{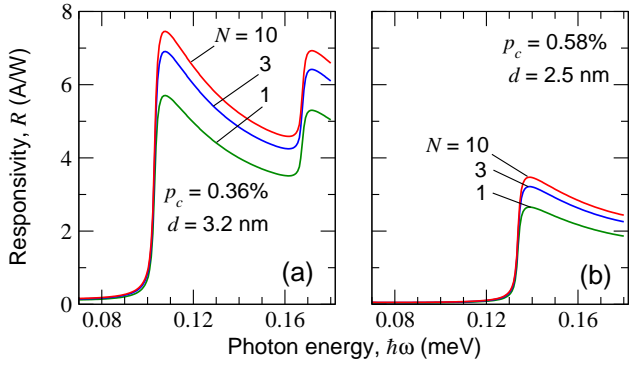}
\caption{ The spectral dependences of the responsivity $R$
for the interband HgTe--Cd$_{0.3}$Hg$_{0.7}$Te QWIPs with different number of the QWs $N$
($N = 1, 3,$ and 10) 
at $T = 200$~K, $\gamma = 0.5$, and $U = E/E^{tunn} = 0.5$: (a)  $d = 3.2$~nm , $p_c = 0.36\%$ and (b)  $d = 2.5$~nm, $p_c =0.58\%$, .}
\end{figure}

Using the above results and considering Eq.~(A1) 
at $\alpha_{\omega} W = \beta_{\omega,PD} < 1$, where $\alpha_{\omega}$ and $W$ are the interband absorption coefficient in the PD
depletion layer and the thickness of the latter, for the ratio  of the pertinent responsivities $R$ and $R_{PD}$ we find

\begin{equation}\label{eq6}
\frac{R}{R_{PD}} \simeq 
\frac{1}{p_c}\frac{\beta_{\omega}}{\beta_{\omega,PD}}. 
\end{equation}
As follows from Eq.~(6),  a relatively small quantum efficiency in the interband QWIPs considered above can be  compensated by  the phoconductive gain (in the interband QWIPs, this gain $g \propto p_c^{-1}$).  For the photon energy  $\hbar\omega \sim 0.1$~eV, assuming $\beta_{\omega} = 3\times 10^{-3}$, $p_c = (0.4 - 0.6)\%$, $\alpha_{\omega} = 2\times 10^3 $~cm$^{-1}$~\cite{31}, and $W = 10^{-4}$~cm  ($\beta_{\omega,PD} = 0.2$), Eq.~(6) yields $R/R_{PD} \simeq 2.5 - 3.75$. 

\subsection{Dark-current-limited detectivity}
\vspace{-0.4cm}

One of the most important figure-of-merit  of the interband QWIP is the dark-current-limited detectivity, which depends on
the detector noise at moderate and elevated temperatures. The detectivity can be expressed via the dark current density and the photoconductive gain.

Compare the interband QWIP detectivity $D^*$ and the p-i-n PD detectivity $D_{PD}^*$ considering
that $D^* \propto \beta_{\omega}N/\sqrt{N G}$ and $D_{PD}^* \propto \beta_{\omega,PD}/\sqrt{2G_{PD}}$,
where $G$ and $G_{PD}$ are the rates of generation in the dark conditions.
We assume that the barrier material in the QWIP and the material of the p-i-n PD are chosen to provide the equal values  of $\hbar\omega_{th} =  \Delta_{QW} + \Delta_B = \Delta_{G}$ [see Eq.~(1)],
where $\Delta_G$ is the energy gap in the depletion region of the p-i-n PDs under comparison.
Since the energy gap in the interband QWIP barrier layers exceeds $ \Delta_{QW} + \Delta_B$, the former is larger than
the energy gap, $\Delta_G$, in the PDs under comparison.
 Due to this, one can neglect the thermogeneration in the barrier layers in comparison to the thermogeneration in the QWs. As a result,

\begin{equation}\label{eq7}
\frac{D^*}{D_{PD}^*} \simeq \frac{\beta_{\omega}N}{\beta_{\omega,PD}}\sqrt{\frac{2G_{PD}}{NG}}.
\end{equation}

Using Eqs.~(9) and (A4), we find

\begin{equation}\label{eq10}
\frac{D^*}{D_{PD}^*} \simeq \frac{\beta_{\omega}}{\beta_{\omega,PD}}\sqrt{\frac{2NW}{\hbar}\sqrt{\frac{mk_BT}{2\pi}}}.
\end{equation}

Assuming $m = 0.025m_0$ ($m_0$ is the mass of bare electron), 
$\beta_{\omega} = 3\times 10^{-3}$, $\alpha_{\omega} = 2\times 10^3 $~cm$^{-1}$, $W = 10^{-4}$~cm (i.e., $\beta_{\omega,PD} = 0.2$),
and $T = 200$~K, 
we obtain from Eq.~(8)
$D^*/D_{PD}\simeq 0.119\sqrt{N}$. Even at $N \sim 10 - 20$, the latter ratio is somewhat smaller than unity.
However,
in reality, the Auger generation  in CdHgTe p-i-n PDs operating in long wavelength radiation range can result in elevated
dark currents and, hence, in a lower detectivity~\cite{32,33} (see also, Refs.~\cite{12,13}.

The QWIPs could exhibit weaker tunneling dark currents providing additional potential advantages of these detectors.

\vspace{-0.4cm}

In conclusion,
we proposed the interband HgTe-CdHgTe QWIPs and analyzed their characteristics.
The analysis of these photodetectors demonstrates their substantial advantages  over the intersubband (intraband)  HgTe-CdHgTe QWIPs, the  conventional p-i-n PDs, and the intersubband  A$_3$B$_5$ QWIPs.

\section*{Acknowledgments}
 \vspace{-0.4cm}

 The work  was supported by   the Russian  Foundation of Basic Research, Grants Nos. 18-52-50024, 18-07-01145,
and  16-29-03033.
 The work at RIEC  and UoA was supported by the Japan Society for Promotion of Science,
KAKENHI Grants Nos. 16H06361 and 16K14243.    
The work at RPI were supported by  the US ARL Cooperative Research Agreement.
VR also acknowledges the support by  the Russian Scientific Foundation, Grants Nos.14-29-00277
and 16-29-03432.

\section*{Appendix. Characteristics of the p-i-n PDs}
\setcounter{equation}{0}
\renewcommand{\theequation} {A\arabic{equation}}

 \vspace{-0.4cm}

Calculating the photocurrent density and the responsivity of the p-i-n PDs with the absorption coefficient $\alpha_{\omega}$
and the thickness of the depletion layer $W$, one can obtain

\begin{equation}\label{eqA1}
 j_{photo, PD} \simeq e\beta_{\omega,PD}I_{\omega}, \qquad R_{PD} \simeq \frac{e\beta_{,\omega,PD}}{\hbar\omega}.
\end{equation}

The volume rates of the thermogeneration, $G$, in one QW and in the depletion region, $G_{PD}$,  can be estimated as

\begin{equation}\label{eqA2}
 G = \frac{mk_BT}{\pi\hbar^2W\tau_R}\exp\biggl(-\frac{\hbar\omega_{th}}{k_BT}\biggr),
\end{equation}

\begin{equation}\label{eqA3}
 G_{PD} = \frac{2(2\pi\,mk_BT)^{3/2}}{(2\pi\hbar)^3\tau_R}\exp\biggl(-\frac{\hbar\omega_{th}}{k_BT}\biggr),
\end{equation}
Here $\tau_R$ is the recombination time (assumed to be equal in both devices).
Equations~(A2) and (A3) yield

\begin{equation}\label{eqA4}
\frac{G_{PD}}{G}  = \sqrt{\frac{mk_BT}{2\pi}}\frac{W}{\hbar}.
\end{equation}
Setting $m = 0.025m_0$, $W = 10^{-4}$~cm, and $T = 200$~K, from Eq.~(A4) we obtain $G_{PD}/G \simeq 31.6$.


\begin{thebibliography}{99}

\bibitem{1} 
K. K. Choi, {\it Physics of Quantum Well Infrared Photodetectors} (World Scientific, Singapore, 1997). 


\bibitem {2}
H. Schneider  and H. C. Lui, {\it Quantum Well Infrared Photodetectors: Physics and Applications} (Springer, Berlin, 2007). 

\bibitem {3}
 A. Rogalski, Quantum well photoconductors in infrared detector technology, J. Appl. Phys. {\bf 93}, 4356 
 (2003). 

\bibitem{4}
F. T. Vasko and A. V. Kuznetsov, {\it Electron States and Optical Transitions in Semiconductor Heterostructures},
(Springer, Berlin 1999).

\bibitem{5}
N. S. Rytova, \lq\lq  Resonance absorption of electromagnetic waves in a thin film, \rq\rq
Sov. Phys. Solid State {\bf 8}, 2136
(1967).

\bibitem{6}
D. D. Coon and R. P. G. Kuranasiri, \lq\lq New mode of IR detection using quantum wells,\rq\rq
Appl. Phys. Lett. {\bf 45}, 649
(1984).


\bibitem{7}
L. C. West and S. J. Eglash, \lq\lq First observation of an exteremely large-dipole infrared transition within the conduction band of a GaAs quantum well,\rq\rq
Appl. Phys. Lett. {\bf 46}, 1156
(1985).


\bibitem {8}
B. F. Levine, 
\lq\lq Quantum‐well infrared photodetectors,\rq\rq 
J. Appl. Phys. {\bf 74}, R1 (1993). 

\bibitem {9}
S. D. Gunapala, S. V. Bandara, J. K. Liu, J.M. Mumolo, D. Z. Ting, C. J. Hill, J. Nguyen , B. Simolon, J. Woolaway, S. C. Wang, W. Li, P. D. LeVan, M. Z. Tidrow, 
\lq\lq 1024 X 1024 Format pixel co-located simultaneously readable dual-band QWIP focal plane,\rq\rq 
J.  Infrared Phys. Technol. {\bf 52},  395
(2009).

\bibitem {10}
S. D. Gunapala, S. V. Bandara, J. K. Liu,
J. M. Mumolo, S. B. Rafol, D. Z. Ting, A. Soibel, and C. Hill, 
\lq\lq Quantum Well Infrared Photodetector Technology and Applications,\rq\rq
IEEE J.  Sel. Topics Quant. Electron. {\bf 20}, No. 6 (2014). 

\bibitem{11}
A. C. Goldberg, St. W. Kennerly, J. W. Little, T. A. Shafer, C. L. Mears, H. F. Schaake, M. L. Winn; M. Taylor; P. N. Uppal, \lq\lq Comparison of HgCdTe and quantum--well infrared photodetector dual-band focal plane arrays,\rq\rq Opt. Eng. {\bf 42}, 1 (2003). 

\bibitem{12}
I. M. Nesmelova and V. A. Andreev, \lq\lq Infrared detector materials alternative to CdHgTe,\rq\rq J. Opt. Technol. {\bf 74}, 217
(2007). 
\bibitem{13}
N. A. Kulakova, A. R. Nasyrov, and I. M. Nesmelova, \lq\lq Current trends in creating optical systems for the IR region,\rq\rq  J. Opt. Technol. {\bf 77}, 324
(2010). 

\bibitem{14}
V. Guériaux,
\lq\lq Quantum well infrared photodetectors: present and future,\rq\rq
 Opt. Eng. {\bf 50}, 061013 (2011).
\bibitem{15}
C. Downs and T. E. Vandervelde, \lq\lq Progress in Infrared Photodetectors Since 2000,\rq\rq
Sensors (Basel) {\bf 13},  5054
(2013).


\bibitem{16}
S. Dvoretsky, N. Mikhailov, Y. Sidorov, V. Shvets, S. Danilov, B. Wittman, and S.D. Ganichev, \lq\lq Growth of HgTe Quantum Wells for IR to THz Detectors,\rq\rq J.  Electronic Materials {\bf 39}, 918 (2010)


\bibitem{17}
S. Morozov, V. Rumyantsev, M. Fadeev, M. Zholudev, K. Kudryavtsev, A.  Antonov, A. Kadykov, A. Dubinov, V. Gavrilenko, N.  Mikhailov, S.  Dvoretsky,  \lq\lq Stimulated emission from HgCdTe quantum well heterostructures at wavelengths up to 19.5$\mu$m,\rq\rq Appl. Phys. Lett. {\bf 111}, 192101  (2017).



\bibitem{18}
S. Ruffenach, A. Kadykov, V. V. Rumyantsev, J. Torres, D. Coquillat, D. But, S. S. Krishtopenko, C. Consejo, W. Knap, S. Winner, M. Helm, M. A. Fadeev, N. N. Mikhailov, S. A. Dvoretzky, V. I. Gavrilenko, S. V. Morozov, and F. Teppe, \lq \lq HgCdTe-based heterostructures for terahertz photonics,\rq\rq
APL  Materials {\bf 5}, 035503 (2017).

\bibitem{19}
Q. Chen, M. Sanderson, and C. Zhang,
\lq\lq Nonlinear terahertz response of HgTe/CdTe quantum wells,\rq\rq Appl. Phys. Lett. {\bf 107}, 081111 (2015).


\bibitem {20}
H. C.  Liu,
\lq\lq Photoconductive gain mechanism of quantum well intersubband infrared detectors,\rq\rq
Appl. Phys. Lett. {\bf 60}, 1507 (1992).

 \bibitem {21}
L. Thibaudeau, P. Bois, and J. Y. Duboz, \lq\lq A self‐consistent model for quantum well infrared photodetectors,\rq\rq J. Appl. Phys. {\bf 79}, 446 (1996). 
 
 \bibitem {22}
 M. Ryzhii, V. Ryzhii, R. Suris, and C. Hamaguchi, 
\lq\lq Self-organization in multiple quantum well infrared photodetectors,\rq\rq
 Semicond. Sci. Technol. {\bf 16}, 202
 (2001).
 
  \bibitem {23}
  V. Ryzhii, I. Khmyrova, M. Ryzhii, R. Suris, and C. Hamaguchi,
  \lq\lq Phenomenological theory of electric-field domains induced infrared radiation in multiple quantum well structures,\rq\rq Phys. Rev. B {\bf 62}, 7268 (2000).

\bibitem {24}
V. Ryzhii, M. Ryzhii, and H. C. Liu,
\lq\lq Self-consistent model for quantum well infrared photodetectors with thermionic injection under dark conditions,\rq\rq
J. Appl. Phys. {\bf 92}, 207 (2002).




\bibitem {25}
V. Ryzhii, M. Ryzhii, D. Svintsov, V. Leiman, V Mitin, M. S. Shur,  and T.  Otsuji,
\lq\lq Infrared photodetectors based on graphene van der Waals heterostructures,\rq\rq
Infrared Phys. Technol. {\bf 84}, 72 (2017).


\bibitem {26}
 V. Ryzhii, M. Ryzhii, D. Svintsov, V. Leiman, V. Mitin, M. S. Shur, and T. Otsuji,
 \lq\lq Nonlinear response of infrared photodetectors based on van der Waals heterostructures with graphene layers,\rq\rq
Optics Express {\bf 25}, 5536 (2017). 

\bibitem {27} 
V. Ryzhii, M. Ryzhii, V. Leiman, V. Mitin, M. S. Shur, and T. Otsuji,
\lq\lq Effect of doping on the characteristics of infrared photodetectors based on van der Waals heterostructures with multiple graphene layer,\rq\rq
J. Appl. Phys. {\bf 122}, 054505 (2017).

\bibitem {28}
S. M. Sze, {\it Physics of Semiconductor Devices}, (Wiley, 1999), p.103.



\bibitem {29}
V. Ya. Aleshkin and A. A, Dubinov, \lq\lq Effect of the spin–orbit interaction on intersubband electron transition in GaAs/AlGaAs quantum well heterostructures,\rq\rq Physica B {\bf 503}, 32 (2016).

\bibitem {30}
V. Ya. Aleshkin, A. A. Dubinov, M. Ryzhii, and V. Ryzhii,
\lq\lq Electron capture in van der Waals graphene-based heterostructures with WS$_2$ barrier layers,\rq\rq
J. Phys. Soc. Japan {\bf 84}, 094703 (2015).

\bibitem {31}
J. H. Chu, B. Li,  K. Liu, and D. Y. Tang, \lq\lq Empirical rule for intrinsic absorption spectroscopy in Hg$_{1-x}$Cd$_x$Te,\rq\rq   J.~Appl. Phys. {\bf 75}, 1234 (1994).



\bibitem {32}
M. Grudzień, K. Jóźwikowski, J. Piotrowski, and H. Polakowski,
\lq The influence of doping on ultimate performance of photodiodes for the 8--14 $\mu$m spectral range,\rq\rq
Infrared Phys.,  {\bf 21}, 201 (1981)


\bibitem {33}
A. Rogalski, \lq\lq HgCdTe infrared detector material: history, status
and outlook,\rq\rq
Rep. Prog. Phys. {\bf 68}, 2267
(2005).


\end{thebibliography}
\end{document}